%% file: structuring_data.tex
\documentclass[conference]{IEEEtran}

\usepackage{comment}
\usepackage{acro}

\usepackage{nohyperref}
\usepackage{graphicx}
\usepackage{siunitx}
\usepackage{tabularray}
\usepackage{comment}
\newcommand{\STAB}[1]{\begin{tabular}{@{}c@{}}#1\end{tabular}}
\usepackage[colorinlistoftodos,prependcaption,textsize=tiny]{todonotes}

\input{acronyms}

\usepackage{tikz}

\newcommand\copyrighttext{%
  \footnotesize \textcopyright \the\year{} IEEE. Personal use of this material is permitted. Permission from IEEE must be obtained for all other uses, including reprinting/republishing this material for advertising or promotional purposes, collecting new collected works for resale or redistribution to servers or lists, or reuse of any copyrighted component of this work in other works.}

\newcommand\copyrightnotice{%
\begin{tikzpicture}[remember picture,overlay]
\node[anchor=south,yshift=10pt] at (current page.south) {\fbox{\parbox{\dimexpr0.75\textwidth-\fboxsep-\fboxrule\relax}{\copyrighttext}}};
\end{tikzpicture}%
}

\title{Structuring Automotive Data for Systems Engineering: A Taxonomy-Based Approach} 
\begin{document}

\author{\IEEEauthorblockN{
Carl Philipp Hohl\IEEEauthorrefmark{1}, 
Philipp Reis\IEEEauthorrefmark{1},
Tobias Schürmann\IEEEauthorrefmark{1},
 Stefan Otten\IEEEauthorrefmark{1},
 and Eric Sax\IEEEauthorrefmark{1}}
\IEEEauthorblockA{\IEEEauthorrefmark{1}FZI Research Center for Information Technology, 76131 Karlsruhe, Germany\\ Email: \{hohl, reis, schuermann, otten, sax\}@fzi.de}}

\IEEEoverridecommandlockouts
\IEEEpubid{\makebox[\columnwidth]{}}

\maketitle

\copyrightnotice
\input{00_Abstract}
\input{01_Introduction}
\input{02_RelatedWorks}
\input{04_methodology}
\input{05_Challenges}


\bibliographystyle{ieeetr}
\bibliography{ref}

\end{document}

%% file: acronyms.tex
\acsetup{
    make-links = false,
}

\DeclareAcronym{ADS}{
    short = ADS,
    long = Automated Driving System,
}

\DeclareAcronym{ADAS}{
    short = ADAS,
    long = Advanced Driving-Assistance System,
}

\DeclareAcronym{VDC}{
    short = VDC,
    long = vehicle data collection,
}

%% file: 00_Abstract.tex
\begin{abstract}
Vehicle data is essential for advancing data-driven development throughout the automotive lifecycle, including requirements engineering, design, verification, and validation, and post-deployment optimization. Developers currently collect data in a decentralized and fragmented manner across simulations, test benches, and real-world driving, resulting in data silos, inconsistent formats, and limited interoperability. This leads to redundant efforts, inefficient integration, and suboptimal use of data. This fragmentation results in data silos, inconsistent storage structures, and limited interoperability, leading to redundant data collection, inefficient integration, and suboptimal application.
To address these challenges, this article presents a structured literature review and develops an inductive taxonomy for automotive data. This taxonomy categorizes data according to its sources and applications, improving data accessibility and utilization. The analysis reveals a growing emphasis on real-world driving and machine learning applications while highlighting a critical gap in data availability for requirements engineering. By providing a systematic framework for structuring automotive data, this research contributes to more efficient data management and improved decision-making in the automotive industry.
\end{abstract}

%% file: 01_Introduction.tex
\section{Introduction}

The trend of automation in the automotive world has led
to a proliferation of data collection and usage. The availability of data aids in requirement elicitation, prototyping, verification, and increasingly the training of AI models. This emergence
in data-driven development methods comes with a host of
new challenges, such as data management, data availability, and data quality assurance. In response, CMMI Institute released their Data Maturity Model (DMM) guidelines, which outline areas and best practices "to help organizations build, improve, and measure their
enterprise data management capability" \cite{cmmiinstituteDataManagementMaturity2019}.
The model assigns maturity levels from 1 to 5 to key areas to improve the generation of value from enterprise data.
This paper addresses the challenges of applying these principles to the automotive sector.
In accordance with the DMM, data governance requires a business glossary, as well as metadata management.
The glossary aims to support a "common understanding of terms and definitions" regarding data and data structure.
A common business language ensures consistent definitions of data-related terms within an organization. At higher maturity levels, organizations must align data models with standardized industry terms.
This structured terminology directly supports metadata management, which according to the DMM helps to establish "processes and infrastructure for specifying and
extending clear and organized information about the structured
and unstructured data assets". 
The use of common terms and wording to denote dataset metadata is necessary in order to enable standardized data capture and retrieval tied to the specific use case.

\begin{figure*}
    \centering
    \includegraphics[width=0.95\textwidth]{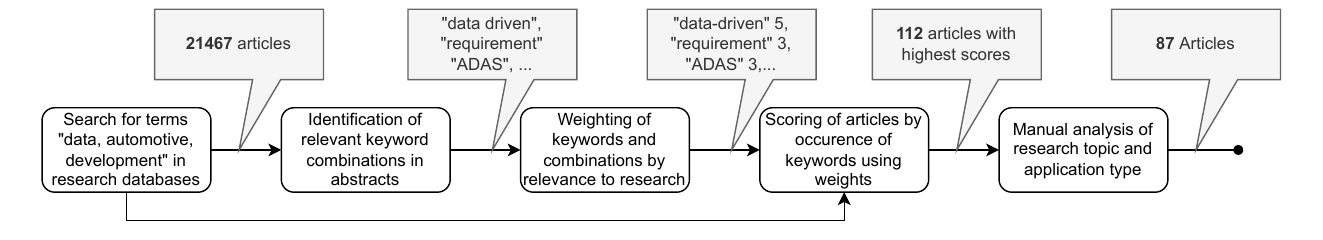}
    \caption{Illustration of our research methodology for reviewing literature leading from 21467 articles to a subset of 87 publications relevant to our research}
    \label{fig:MethPaper}
\end{figure*}

%% file: 02_RelatedWorks.tex
\section{Related Work}

The utilization of vehicle data has been a topic of growing interest in recent years, with various approaches emerging to harness its potential. To provide a comprehensive understanding, we examine the different perspectives on vehicle data utilization. Since there are differing viewpoints and applications for data-driven methods and as a result, the way in which data is categorized and viewed, multiple models have emerged to provide better understanding of automotive data. 

\paragraph{Ontology-based Perspective}
An ontology-based data model to enhance analytics and machine learning capabilities in the automotive industry is proposed in~\cite{ALVAREZCOELLO2021}. This work addresses the need for standardized data models in a data-centric architecture, facilitating the shift towards data-driven functionality. For instance, the authors highlight the importance of standardizing vehicle signal specifications to enable seamless communication between different components. However, this paper does not systematically analyze the data itself nor specify the exact purpose of data usage.

\paragraph{Signal-based Perspective}
A signal-based perspective is presented as Vehicle Signal Specification (VSS)~\cite{EURECOM_2018}  and extended to an ontology in ~\cite{Wilms2021}. This ontology describes the internal connections within the vehicle signal network and its domain, providing a detailed understanding of the relationships between different signals. Nevertheless, this approach is limited to vehicle signals, excluding sensor modalities like cameras or LiDAR.

\paragraph{Process Perspective}
A data-driven development approach is proposed by Bach et al.~\cite{bach_data-driven_2017}, showcasing the potential value of data in all vehicle phases through an application of predictive cruise control. This work briefly describes data usage across various development phases and concretely illustrates it with a single use case, demonstrating the feasibility of integrating data-driven approaches into existing development processes.

\paragraph{Scenario Perspective}
Zhang et al.~\cite{zhang_finding_2021} describe the identification of critical scenarios in a taxonomic manner, highlighting its potential support for data-driven development in automated driving. However, this contribution focuses on critical scenarios, covering only a small portion of all vehicle data.

\paragraph{Data-centric Perspective}
A data-centric perspective describing the importance of real-world test drive and customer drive data in automotive system development by proposing a 4W1H framework~\cite{Sohn24}. This answers the question of why, where, what, and how data should be collected from real-world test drives. While this proposal gives insight into a data-centric perspective for the use of real-world vehicle data, it is limited to an operational design domain view. 
\\
These approaches differ with respect to the problem they try to address. While~\cite{ALVAREZCOELLO2021},~\cite{EURECOM_2018}, and~\cite{Wilms2021} focus on understanding at the system level different data types and aims to improve standardization, \cite{bach_data-driven_2017} and ~\cite{Sohn24} focus more on a process point of view for value generation utilizing the systematic collection and usage of data.
\\

\section{Contribution}

\begin{figure*}
    \centering
    \includegraphics[width=\textwidth]{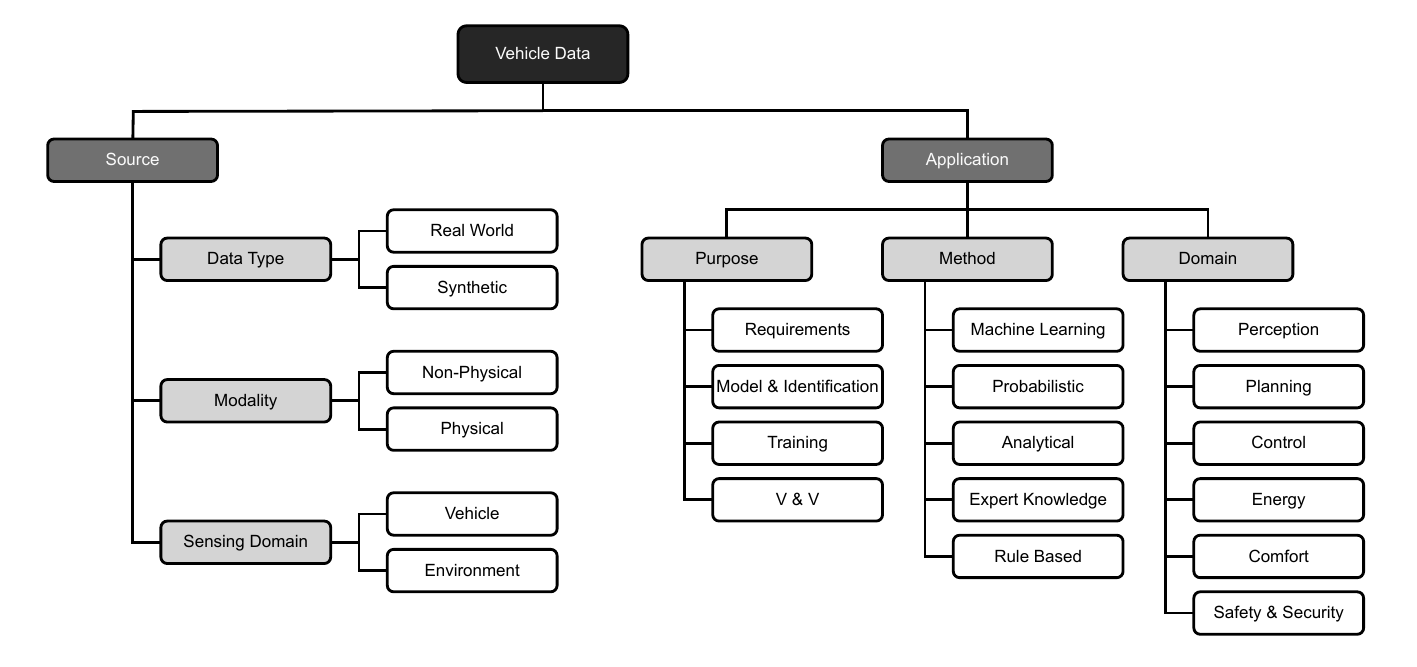}    
    \caption{Proposed taxonomy for vehicle real-world data collection derived from the comprehensive literature review illustrated in Fig.~\ref{fig:MethPaper}}
    \label{fig:MapKeyDriver}
\end{figure*}

Our contribution shares a data-centric perspective, presenting the current state in the form of a taxonomy. The proposed taxonomy aims to systematically classify vehicle data, offering a structured framework to address key challenges in automotive system development.

This study addresses the need for a unified language in the field of vehicle data use
by providing a comprehensive review of the existing literature
and proposing an extensible vehicle data taxonomy. We aim to
classify different types of data, that are generated within vehicle architectures based on abstract properties. These
are informed by the point of origin of the data and the potential use to different stakeholders
and should allow other views to be introduced without
contradiction while providing added value to businesses, developers, and
researchers. The proposed framework categorizes data from
various sources (e.g. sensors, simulations, and real-world operations) into distinct categories, offering a structured approach
to understanding the interrelationships between vehicle data
collection, integration, and application. We work towards
closing the gap between viewing data from a syntactic point
of view focusing on the application within the automotive
development process by adding classifications that inform
about the origin, semantics, and abstraction level associated
with data used in development and research. Additionally, our
study examines the purpose, methods, and domain of vehicle
data usage, providing actionable insights for optimizing data
management capabilities and improve organizational maturity. Using our proposed taxonomy, stakeholders can inform
decision-making in data collection strategies, usage planning,
and management processes. For example, manufacturers can
utilize our taxonomy to develop targeted data collection strategies, while researchers can apply our findings to advance data-
driven methodologies in vehicle development.

\subsection{Research Questions}
In light of the growing application of data in the automotive sector and availability of more data, in companies as well as in the research field, we ask the following questions:

\begin{itemize}
    \item \emph{Research Question A: What different types of automotive data are collected and how can it be classified?} 
    \item \emph{Research Question B: What types of application do these data have and how can a classification be used within the context of a data model?}
\end{itemize}

%% file: 04_methodology.tex
\subsection{Methodology}

In order to answer \textit{Research Question A}, we conducted an analysis of applications, which are presently employing data in any activity related to the automotive development. To this end, we compiled 21467 research articles published since 2015 by searching for the term "data, automotive, development" in IEEE (9994 results) and Scopus (11473 results). We then conducted an analysis of keywords and their interrelation within the abstracts of those papers and their relation with each other (see Figure~\ref{fig:MethPaper}). The result of this analysis is a word cloud of interrelated terms that are used within this field of research. We identified significant keywords that are most often used to describe data based activities during development and life cycle and rated them by their relevance to our research questions. Finally, we graded each article with regards to the occurrence of those keywords and sorted them by relevance. This allowed us to exclude search terms that hint towards applications that are not in our area of interest and perform a weighting on terms that are more relevant. This offered an improved ranking compared to using the search functionality of research publishers.\\
We used the following weights (case insensitive):
\begin{itemize}
\item 5: \textit{data-driven, driving data, real-world data}
\item 4: \textit{data collection, data selection, data aquisition, vehicle, car, SAE Level, SAE LVL}
\item 3: \textit{verification, validation, requirements, design, architecture, ADAS, AD, autonomous, testing, test, camera, lidar, CAN, radar}
\item 2: \textit{development}
\end{itemize}
We excluded papers using the following terms: \\
\textit{UAV, smartphone, unmanned, attack, marine, AUV, underwater, train} \\
We only considered papers published from 2015 up until 2024 and required a minimum average of five citations per year.
Using the 112 most relevant papers, we sorted them into broad categories, such as \textit{Development}, \textit{Machine Learning Application} and \textit{Service provision}.  We then refined our categories by separating the research into categories of application and derived a taxonomy of data application in the automotive research sector. After manual elimination of papers unsuited towards our taxonomy, we assigned the remaining 87 papers to the categories (see Table~\ref{tab:TaxTable}). Papers that were omitted include topic mismatches such as different fields of application or papers in which no source or application for data.

\section{Overview of Taxonomy}

\begin{table*}

\caption{Literature concerning taxonomic elements}
\centering
\small 
\begin{tblr}{
  colspec={Q[c]Q[c]Q[l]Q[l]}, 
  cell{1}{1} = {r=6}{},
  cell{7}{1} = {r=15}{},
  cell{1}{2} = {r=2}{},
  cell{3}{2} = {r=2}{},
  cell{5}{2} = {r=2}{},
  cell{7}{2} = {r=4}{},
  cell{11}{2} = {r=5}{},
  cell{16}{2} = {r=6}{},
  column{4}={0.75\linewidth},
  hlines, 
  vlines, 
  hline{1,8,23} = {-}{}, 
  hline{2-3,5,7} = {3-4}{}, 
  hline{4,6,12,17} = {3-4}{},
}
\label{tab:TaxTable}
\STAB{\rotatebox[origin=c]{90}{Source}} 
    & \STAB{\rotatebox[origin=c]{90}{Data Type}} 
        & Real World & \input{chapter/table_cites/source/real_world} \\
    & & Synthetic & \input{chapter/table_cites/source/synth} \\
    
    & \STAB{\rotatebox[origin=c]{90}{Modality}} 
        & Physical & \input{chapter/table_cites/source/physical}\\
    & & Non-Physical & \input{chapter/table_cites/source/nonPhys} \\
    & \STAB{\rotatebox[origin=c]{90}{Sensing Domain}} 
        & Vehicle & \input{chapter/table_cites/source/vehicle} \\
    & & Environment & \input{chapter/table_cites/source/environment} \\
\STAB{\rotatebox[origin=c]{90}{Application}} 
    & \STAB{\rotatebox[origin=c]{90}{Purpose}} 
        & Requirements & \input{chapter/table_cites/purpose/requirement} \\
    & & Model \& Identification & \input{chapter/table_cites/purpose/mi} \\
    & & Training & \input{chapter/table_cites/purpose/train} \\
    & & V \& V & \input{chapter/table_cites/purpose/vAndV} \\
    
    & \STAB{\rotatebox[origin=c]{90}{Methods}} 
        & Machine Learning & \input{chapter/table_cites/methods/ml} \\
    & & Probabilistic & \input{chapter/table_cites/methods/prob} \\
    & & Analytical & \input{chapter/table_cites/methods/analytic} \\
    & & Expert Knowledge & \input{chapter/table_cites/methods/expert} \\
    & & Rule Based & \input{chapter/table_cites/methods/rule} \\

    & \STAB{\rotatebox[origin=c]{90}{Domain}} 
        & Perception & \input{chapter/table_cites/domain/perception} \\
    & & Planning & \input{chapter/table_cites/domain/planning} \\
    & & Control & \input{chapter/table_cites/domain/control} \\
    & & Safety \& Security & \input{chapter/table_cites/domain/safety} \\
    & & Comfort & \input{chapter/table_cites/domain/comfort} \\
    & & Energy & \input{chapter/table_cites/domain/energy} \\
\end{tblr}
\end{table*}

After an analysis of the literature, we divided the taxonomy into two broad groups of data source and data application (see Fig.~\ref{fig:MapKeyDriver}). The data source specifies where the data that is being used within the paper originates. In order to further specify the source, we defined the following categories: \\
\subsubsection{Source}
\textbf{Data Type} specifies how the data was obtained. This category can inform us about the realism and availability of data during research activities. \textbf{Real World} data describes data that has been sampled from real world systems, such as real sensor data obtained from a test drive and recording the sensor data. \textbf{Synthetic} data originates from simulators that are used to generate synthetic signals, mostly from models that describe sensor or system behavior or has been generated through other means than simulation. This mostly encompasses machine learning models or style-transfer methods as well as augmentation of other data.\\
\textbf{Modality} or abstraction of the data informs us of the nature of the data, specifically if it refers to physically observable phenomena. \textbf{Physical} data relates to physical phenomena that can be measured using sensors. Examples of this are accelerations, angles, pressures or positions. Also within this category fall camera images, as they record the light values reaching a sensor. With regards to the chain of control within a system, they often are input or output data, since they have a direct relation to the real world. 
\textbf{Non-Physical} data is calculated data that is conceptual in nature and is not directly observable as a physical phenomenon. This could be categorical data such as classes of objects within an image, trajectory data describing the movement of a vehicle over time or an measure of a drivers alertness. This data can not be obtained by a sensor directly but is always the result of some kind of computation. It is often an internal or intermediary value within a control chain.
The distinction between physically based data (e.g., sensor readings) and non-physically based data (e.g., behavior patterns) matters as it impacts validation, application context, and analysis methods. Understanding this distinction helps address data gaps and ensure accurate and effective use in automotive research.
The \textbf{Sensing Domain} describes the domain within which the data holds meaning within the frame of automotive data. As the system under consideration is a vehicle, the data can either describe a state of the vehicle or a state of the environment. Thus, \textbf{Vehicle} data is all data related to the vehicle itself. This can be its position, velocity or other internal state of the car. \textbf{Environment} data is all data measured outside the vehicle. This can include other agents or data received from third parties via external interfaces not pertaining to the car itself. \\

\subsubsection{Application}

Within the \textbf{Application} category we distinguish between three categories related to the usage of the data within the literature. \\ The \textbf{Application Purpose} describes during what part of the development or life cycle the data is utilized. We identified \textbf{Requirements}, describing all activities relating to requirements engineering and elicitation. An article is added to this category, if data was used to find, improve or validate requirements related to a system. 
\textbf{Model \& Identification} encompasses the usage of data for use in modeling or identifying parameters of a system or subsystem. This can be, for example, the tuning of a control system or modelling the behaviour of agents for a simulation or conflict resolution.
\textbf{Training} describes the usage of data for training machine learning or statistical models, either for productive use or for other activities.
\textbf{V \& V} is the activity of verification and validation of a system or subsystem. We included all usage of data to verify or validate a systems behavior. Data is used to stimulate the system and outputs of the system are either compared with requirements or validated to produce the correct behavior.

The application \textbf{Methods} section categorizes the application by the methods used within the article. These vary widely but we established the following categories.
\textbf{Machine Learning} methods are all methods using machine learning or similar systems. We included all methods that use data in training a model to produce a certain output.
\textbf{Probabilistic} methods use probabilities or statistics to gain knowledge or insights from data. \textbf{Analytical} methods is the application of an algorithm or calculation in order to model or calculate an output via analytic formulas as opposed to statistical means.
\textbf{Expert Knowledge} describes the application of expert knowledge to the data used. Examples of this are manual comparison of scenarios \cite{stadlerCredibilityAssessmentApproach2022} or optimization with the use of expert opinions \cite{huang_multitarget_2023}.
\textbf{Rule Based} methods are as the name implies tasks that rely on classification or processing based on rules applied to the data.\\
Application \textbf{Domain} is the system domain within which the data is used. Since automation is a major trend in the automotive sector at the time, many articles refer to ADAS/AD applications, which are either Advanced Driver Assistance Systems or Autonomous Driving. Since this category is very broad, we further subdivided it into the activities of Perception, Planning and Control. For the sake of simplicity we use these three categories for all automated driving usage of data. 
\textbf{Perception} is all data use related to perceiving the environment, such as object and lane detection or intent recognition.
\textbf{Planning} refers to all planning related data usage, such as trajectory planning, environment modeling or optimization.
\textbf{Control} applications are involved in the control of subsystems, such as velocity, steering or other control tasks.
\textbf{Safety \& Security} relates to all usage of data in order to estimate or improve the security and safety of a system. This can describe tasks such as predictive maintenance or driver attention estimation.
\textbf{Comfort} are all activities related to the comfort of the passengers. This can include the improvement of air conditioning or personalized functions.
\textbf{Energy} relates to all energy management tasks. This can be vehicle based or larger scale and includes tasks such as charge forecasting, route optimization or energy conservation.

%% file: chapter/table_cites/source/real_world.tex
\cite{gao19,shi_memory-based_2022,bedretchuk_low-cost_2023,moreira-matias_developing_2017,Deo_looking_2018,pham3DDatasetAutonomous2020,schmidDataDrivenFaultDiagnosis2021,xingChargingDemandForecasting2019,xingDynamicStateEstimation2020,zhaoAcceleratedEvaluationAutomated2018,aminiCabinBatteryThermal2020,birchlerCosteffectiveSimulationbasedTest2022,degelderScenarioParameterGeneration2022,deoLookingDriverRider2020,fowlerFuzzTestingAutomotive2018,gaoPersonalizedAdaptiveCruise2020,gwonGenerationPreciseEfficient2017,hanResearchRoadEnvironmental2023,jaguemontCharacterizationModelingHybridElectricVehicle2016,kimTakeoverRequestsSimulated2017,kongFedVCPFederatedLearningBasedCooperative2022,lvCharacterizationDriverNeuromuscular2018,lyuUsingNaturalisticDriving2022,rahmatiHelpingAutomatedVehicles2022,shinVehicleSpeedPrediction2019,stadlerCredibilityAssessmentApproach2022,tianDeepTestAutomatedTesting2018,wangBChargeDataDrivenRealTime2018,wangVibrationTheoreticApproachVulnerability2023,xiaoTrajDataVehicleTrajectory2020,santaLPWANBasedVehicularMonitoring2019,leeDesignImplementationEdgeFogCloud2020,shiConstructingFundamentalDiagram2021,shinComparativeStudyMarkov2021,jaguemontCharacterizationModelingHybridElectricVehicle2016a,leledakisMethodPredictingCrash2021,choiTrajGAILGeneratingUrban2021,CredibilityAssessmentApproach,benkraoudaTrafficDataImputation2020,arvinRolePrecrashDriving2019,caiIntegrationGPSMonocular2018,ILOCuSIncentivizingVehicle,waliHowDrivingVolatility2018,liuDatadrivenEnergyManagement2021,liuCustomizingDrivingCycles2016,xingChargingDemandForecasting2019a,zhaoTrafficNetOpenNaturalistic2017,zhaoElectricVehicleBatteries2022,vicenteLinearSystemIdentification2021,loniDatadrivenEquitablePlacement2023,haqueLoRaArchitectureV2X2020,fridmanMITAdvancedVehicle2019b}

%% file: chapter/table_cites/source/synth.tex
\cite{zhaoAcceleratedEvaluationAutomated2018, birchlerCosteffectiveSimulationbasedTest2022, leledakisMethodPredictingCrash2021, CredibilityAssessmentApproach, caiIntegrationGPSMonocular2018,  chenAutonomousVehicleTesting2018,  chenDeepImitationLearning2019,  chenRHONNModellingEnabledNonlinear2022,  gidadoSurveyDeepLearning2020,  luLearningDriverSpecificBehavior2018,  qiLearningBasedPathPlanning2023,  tangATACBasedCarFollowingModel2022,  wangSFNetNImprovedSFNet2022,  weiContractBasedChargingProtocol2023,  xuGenerativeAIEmpoweredSimulation2023, wangDataDrivenPredictiveControl2022, nieDeepNeuralNetworkBasedModellingLongitudinalLateral2022, cuiCoopernautEndtoEndDriving2022, tangDriverLaneChange2020a}

%% file: chapter/table_cites/source/physical.tex
\cite{gao19, shi_memory-based_2022, moreira-matias_developing_2017,  pham3DDatasetAutonomous2020,  schmidDataDrivenFaultDiagnosis2021,  xingDynamicStateEstimation2020,  deoLookingDriverRider2020,  gaoPersonalizedAdaptiveCruise2020,  gwonGenerationPreciseEfficient2017,  hanResearchRoadEnvironmental2023,  kongFedVCPFederatedLearningBasedCooperative2022,  lvCharacterizationDriverNeuromuscular2018,  lyuUsingNaturalisticDriving2022,  rahmatiHelpingAutomatedVehicles2022,  shinVehicleSpeedPrediction2019,  xiaoTrajDataVehicleTrajectory2020, leeDesignImplementationEdgeFogCloud2020, jaguemontCharacterizationModelingHybridElectricVehicle2016a, leledakisMethodPredictingCrash2021, CredibilityAssessmentApproach, waliHowDrivingVolatility2018, liuDatadrivenEnergyManagement2021, liuCustomizingDrivingCycles2016, chenDeepImitationLearning2019,  chenRHONNModellingEnabledNonlinear2022,  gidadoSurveyDeepLearning2020,  luLearningDriverSpecificBehavior2018,  qiLearningBasedPathPlanning2023,  wangSFNetNImprovedSFNet2022,  weiContractBasedChargingProtocol2023,  xuGenerativeAIEmpoweredSimulation2023, wangDataDrivenPredictiveControl2022, nieDeepNeuralNetworkBasedModellingLongitudinalLateral2022, cuiCoopernautEndtoEndDriving2022, berziDevelopmentDrivingCycles2016a}

%% file: chapter/table_cites/source/nonPhys.tex
 \cite{shi_memory-based_2022, moreira-matias_developing_2017,  xingChargingDemandForecasting2019,  zhaoAcceleratedEvaluationAutomated2018, aminiCabinBatteryThermal2020,  birchlerCosteffectiveSimulationbasedTest2022,  degelderScenarioParameterGeneration2022,  fowlerFuzzTestingAutomotive2018,  kimTakeoverRequestsSimulated2017,  stadlerCredibilityAssessmentApproach2022,  tianDeepTestAutomatedTesting2018,  wangVibrationTheoreticApproachVulnerability2023, shiConstructingFundamentalDiagram2021, choiTrajGAILGeneratingUrban2021, benkraoudaTrafficDataImputation2020, arvinRolePrecrashDriving2019, xingChargingDemandForecasting2019a, fridmanMITAdvancedVehicle2019b,  tangATACBasedCarFollowingModel2022, berziDevelopmentDrivingCycles2016a, weiEfficientDatadrivenOptimal2022, ebelForcesDriverDistraction2023a}

%% file: chapter/table_cites/source/vehicle.tex
\cite{bedretchuk_low-cost_2023,  pham3DDatasetAutonomous2020,  xingChargingDemandForecasting2019, aminiCabinBatteryThermal2020,  degelderScenarioParameterGeneration2022,  gwonGenerationPreciseEfficient2017,  hanResearchRoadEnvironmental2023,  rahmatiHelpingAutomatedVehicles2022,  wangVibrationTheoreticApproachVulnerability2023, benkraoudaTrafficDataImputation2020, liuDatadrivenEnergyManagement2021, liuCustomizingDrivingCycles2016, xingChargingDemandForecasting2019a, vicenteLinearSystemIdentification2021, fridmanMITAdvancedVehicle2019b,  chenDeepImitationLearning2019,  tangATACBasedCarFollowingModel2022,  wangSFNetNImprovedSFNet2022,  xuGenerativeAIEmpoweredSimulation2023, wangDataDrivenPredictiveControl2022, nieDeepNeuralNetworkBasedModellingLongitudinalLateral2022, berziDevelopmentDrivingCycles2016a, fengDesignDistributedCyber2018}

%% file: chapter/table_cites/source/environment.tex
\cite{gao19, shi_memory-based_2022, moreira-matias_developing_2017, Deo_looking_2018,  schmidDataDrivenFaultDiagnosis2021,  xingDynamicStateEstimation2020,  zhaoAcceleratedEvaluationAutomated2018, aminiCabinBatteryThermal2020,  birchlerCosteffectiveSimulationbasedTest2022,  deoLookingDriverRider2020,  fowlerFuzzTestingAutomotive2018,  gaoPersonalizedAdaptiveCruise2020,  kimTakeoverRequestsSimulated2017,  kongFedVCPFederatedLearningBasedCooperative2022,  lvCharacterizationDriverNeuromuscular2018,  lyuUsingNaturalisticDriving2022,  shinVehicleSpeedPrediction2019,  tianDeepTestAutomatedTesting2018,  xiaoTrajDataVehicleTrajectory2020, leeDesignImplementationEdgeFogCloud2020, shinComparativeStudyMarkov2021, leledakisMethodPredictingCrash2021, CredibilityAssessmentApproach, arvinRolePrecrashDriving2019, caiIntegrationGPSMonocular2018, haqueLoRaArchitectureV2X2020, fridmanMITAdvancedVehicle2019b,  chenDeepImitationLearning2019,  chenRHONNModellingEnabledNonlinear2022,  gidadoSurveyDeepLearning2020,  luLearningDriverSpecificBehavior2018,  qiLearningBasedPathPlanning2023,  weiContractBasedChargingProtocol2023, wangDataDrivenPredictiveControl2022, cuiCoopernautEndtoEndDriving2022, tangDriverLaneChange2020a, ebelForcesDriverDistraction2023a, deyVehicletovehicleV2VVehicletoinfrastructure2016}

%% file: chapter/table_cites/purpose/requirement.tex
\cite{birchlerCosteffectiveSimulationbasedTest2022,  degelderScenarioParameterGeneration2022,  wangBChargeDataDrivenRealTime2018, berziDevelopmentDrivingCycles2016a, weiEfficientDatadrivenOptimal2022, moreira-matiasDevelopingDriverIdentification2017}

%% file: chapter/table_cites/purpose/mi.tex
\cite{schmidDataDrivenFaultDiagnosis2021,  xingChargingDemandForecasting2019,  gaoPersonalizedAdaptiveCruise2020,  gwonGenerationPreciseEfficient2017,  jaguemontCharacterizationModelingHybridElectricVehicle2016,  kimTakeoverRequestsSimulated2017,  lvCharacterizationDriverNeuromuscular2018,  lyuUsingNaturalisticDriving2022,  rahmatiHelpingAutomatedVehicles2022,  shinVehicleSpeedPrediction2019,  wangBChargeDataDrivenRealTime2018,  wangVibrationTheoreticApproachVulnerability2023,  xiaoTrajDataVehicleTrajectory2020, jaguemontCharacterizationModelingHybridElectricVehicle2016a, leledakisMethodPredictingCrash2021, caiIntegrationGPSMonocular2018, liuCustomizingDrivingCycles2016, xingChargingDemandForecasting2019a, vicenteLinearSystemIdentification2021, chenRHONNModellingEnabledNonlinear2022,  tangATACBasedCarFollowingModel2022} 

%% file: chapter/table_cites/purpose/train.tex
\cite{moreira-matias_developing_2017, Deo_looking_2018,  pham3DDatasetAutonomous2020,  xingDynamicStateEstimation2020,  deoLookingDriverRider2020,  kongFedVCPFederatedLearningBasedCooperative2022, shinComparativeStudyMarkov2021, choiTrajGAILGeneratingUrban2021, CredibilityAssessmentApproach, benkraoudaTrafficDataImputation2020, zhaoElectricVehicleBatteries2022, chenDeepImitationLearning2019,  gidadoSurveyDeepLearning2020,  luLearningDriverSpecificBehavior2018,  qiLearningBasedPathPlanning2023,  tangATACBasedCarFollowingModel2022,  wangSFNetNImprovedSFNet2022, nieDeepNeuralNetworkBasedModellingLongitudinalLateral2022, tangDriverLaneChange2020a, ebelForcesDriverDistraction2023a, neurohrCriticalityAnalysisVerification2021}

%% file: chapter/table_cites/purpose/vAndV.tex
\cite{gao19, bedretchuk_low-cost_2023,  zhaoAcceleratedEvaluationAutomated2018, aminiCabinBatteryThermal2020,  birchlerCosteffectiveSimulationbasedTest2022,  degelderScenarioParameterGeneration2022,  fowlerFuzzTestingAutomotive2018,  hanResearchRoadEnvironmental2023,  jaguemontCharacterizationModelingHybridElectricVehicle2016,  rahmatiHelpingAutomatedVehicles2022,  stadlerCredibilityAssessmentApproach2022,  tianDeepTestAutomatedTesting2018,  chenAutonomousVehicleTesting2018,  weiContractBasedChargingProtocol2023,  xuGenerativeAIEmpoweredSimulation2023}

%% file: chapter/table_cites/methods/ml.tex
\cite{moreira-matias_developing_2017, Deo_looking_2018,  xingChargingDemandForecasting2019,  xingDynamicStateEstimation2020, birchlerCosteffectiveSimulationbasedTest2022,  deoLookingDriverRider2020,  kongFedVCPFederatedLearningBasedCooperative2022,  lyuUsingNaturalisticDriving2022,  tianDeepTestAutomatedTesting2018,  shinComparativeStudyMarkov2021,  choiTrajGAILGeneratingUrban2021,  benkraoudaTrafficDataImputation2020,  liuDatadrivenEnergyManagement2021,  liuCustomizingDrivingCycles2016,  fridmanMITAdvancedVehicle2019b,  chenDeepImitationLearning2019,  chenRHONNModellingEnabledNonlinear2022,  gidadoSurveyDeepLearning2020,  luLearningDriverSpecificBehavior2018,  qiLearningBasedPathPlanning2023,  tangATACBasedCarFollowingModel2022,  wangSFNetNImprovedSFNet2022,  xuGenerativeAIEmpoweredSimulation2023,  wangDataDrivenPredictiveControl2022, cuiCoopernautEndtoEndDriving2022,  tangDriverLaneChange2020a,  weiEfficientDatadrivenOptimal2022,  ebelForcesDriverDistraction2023a,  moreira-matiasDevelopingDriverIdentification2017,  neurohrCriticalityAnalysisVerification2021, aliMachineLearningTechnologies2021,  bedretchukLowCostDataAcquisition2023,  DesignTrafficEmergency,  LookingDriverRider,  ostadianIntelligentEnergyManagement2020a, zouVehicleAccelerationPrediction2022}

%% file: chapter/table_cites/methods/prob.tex
\cite{schmidDataDrivenFaultDiagnosis2021,  zhaoAcceleratedEvaluationAutomated2018, degelderScenarioParameterGeneration2022,  fowlerFuzzTestingAutomotive2018,  shinVehicleSpeedPrediction2019,  wangBChargeDataDrivenRealTime2018,  shinComparativeStudyMarkov2021, arvinRolePrecrashDriving2019,  caiIntegrationGPSMonocular2018,  waliHowDrivingVolatility2018,  liuDatadrivenEnergyManagement2021, aliMachineLearningTechnologies2021,  zouVehicleAccelerationPrediction2022}

%% file: chapter/table_cites/methods/analytic.tex
\cite{gao19, shi_memory-based_2022, aminiCabinBatteryThermal2020,  gaoPersonalizedAdaptiveCruise2020,  gwonGenerationPreciseEfficient2017,  hanResearchRoadEnvironmental2023,  jaguemontCharacterizationModelingHybridElectricVehicle2016,  kimTakeoverRequestsSimulated2017,  lvCharacterizationDriverNeuromuscular2018,  wangBChargeDataDrivenRealTime2018,  wangVibrationTheoreticApproachVulnerability2023,  xiaoTrajDataVehicleTrajectory2020, liuDatadrivenEnergyManagement2021, xingChargingDemandForecasting2019a,  vicenteLinearSystemIdentification2021,  chenAutonomousVehicleTesting2018,  weiContractBasedChargingProtocol2023, wangDataDrivenPredictiveControl2022,  neurohrCriticalityAnalysisVerification2021,  zouVehicleAccelerationPrediction2022, huangMultitargetPredictionOptimization2023b,  jiaPlatoonBasedCooperative2016,  qiTrafficDifferentiatedClustering2020, gaoMultisensorFusionRoad2019}

%% file: chapter/table_cites/methods/expert.tex
\cite{jaguemontCharacterizationModelingHybridElectricVehicle2016,  stadlerCredibilityAssessmentApproach2022, neurohrCriticalityAnalysisVerification2021, huangMultitargetPredictionOptimization2023b}

%% file: chapter/table_cites/methods/rule.tex
\cite{schmidDataDrivenFaultDiagnosis2021, fowlerFuzzTestingAutomotive2018,  kimTakeoverRequestsSimulated2017,  rahmatiHelpingAutomatedVehicles2022,  weiContractBasedChargingProtocol2023}

%% file: chapter/table_cites/domain/perception.tex
\cite{pham3DDatasetAutonomous2020, birchlerCosteffectiveSimulationbasedTest2022,  deoLookingDriverRider2020,  gwonGenerationPreciseEfficient2017,  hanResearchRoadEnvironmental2023,  lvCharacterizationDriverNeuromuscular2018,  lyuUsingNaturalisticDriving2022,  stadlerCredibilityAssessmentApproach2022,  chenDeepImitationLearning2019,  wangSFNetNImprovedSFNet2022,  xuGenerativeAIEmpoweredSimulation2023}

%% file: chapter/table_cites/domain/planning.tex
\cite{xingDynamicStateEstimation2020,  degelderScenarioParameterGeneration2022,  kimTakeoverRequestsSimulated2017,  kongFedVCPFederatedLearningBasedCooperative2022,  rahmatiHelpingAutomatedVehicles2022,  xiaoTrajDataVehicleTrajectory2020, chenAutonomousVehicleTesting2018,  gidadoSurveyDeepLearning2020,  luLearningDriverSpecificBehavior2018,  qiLearningBasedPathPlanning2023,  tangATACBasedCarFollowingModel2022}

%% file: chapter/table_cites/domain/control.tex
\cite{gao19,  gaoPersonalizedAdaptiveCruise2020,  hanResearchRoadEnvironmental2023,  jaguemontCharacterizationModelingHybridElectricVehicle2016,  lvCharacterizationDriverNeuromuscular2018,  shinVehicleSpeedPrediction2019,  wangVibrationTheoreticApproachVulnerability2023,  xiaoTrajDataVehicleTrajectory2020,  vicenteLinearSystemIdentification2021, chenRHONNModellingEnabledNonlinear2022,  qiLearningBasedPathPlanning2023,  wangSFNetNImprovedSFNet2022,  wangDataDrivenPredictiveControl2022,  nieDeepNeuralNetworkBasedModellingLongitudinalLateral2022, jiaPlatoonBasedCooperative2016,  yaoDynamicPredictiveTraffic2020}

%% file: chapter/table_cites/domain/safety.tex
\cite{moreira-matias_developing_2017, Deo_looking_2018,  zhaoAcceleratedEvaluationAutomated2018, birchlerCosteffectiveSimulationbasedTest2022,  degelderScenarioParameterGeneration2022,  deoLookingDriverRider2020,  fowlerFuzzTestingAutomotive2018,  kimTakeoverRequestsSimulated2017,  lyuUsingNaturalisticDriving2022,  stadlerCredibilityAssessmentApproach2022,  tianDeepTestAutomatedTesting2018, arvinRolePrecrashDriving2019,  waliHowDrivingVolatility2018,  chenAutonomousVehicleTesting2018,  xuGenerativeAIEmpoweredSimulation2023,  ebelForcesDriverDistraction2023a,  neurohrCriticalityAnalysisVerification2021, aliMachineLearningTechnologies2021,  DesignTrafficEmergency,  LookingDriverRider,  zouVehicleAccelerationPrediction2022, almeaibedDigitalTwinAnalysis2021,  cebeBlock4ForensicIntegratedLightweight2018,  parkScenarioMiningLevel42021a}

%% file: chapter/table_cites/domain/comfort.tex
\cite{aminiCabinBatteryThermal2020,  gaoPersonalizedAdaptiveCruise2020,  kimTakeoverRequestsSimulated2017,  luLearningDriverSpecificBehavior2018, huangMultitargetPredictionOptimization2023b}

%% file: chapter/table_cites/domain/energy.tex
\cite{shi_memory-based_2022,  schmidDataDrivenFaultDiagnosis2021,  xingChargingDemandForecasting2019, aminiCabinBatteryThermal2020,  jaguemontCharacterizationModelingHybridElectricVehicle2016,  lyuUsingNaturalisticDriving2022,  wangBChargeDataDrivenRealTime2018,  jaguemontCharacterizationModelingHybridElectricVehicle2016a,  liuDatadrivenEnergyManagement2021,  liuCustomizingDrivingCycles2016,  zhaoElectricVehicleBatteries2022,  loniDatadrivenEquitablePlacement2023,  weiContractBasedChargingProtocol2023, berziDevelopmentDrivingCycles2016a,  weiEfficientDatadrivenOptimal2022,  ostadianIntelligentEnergyManagement2020a}

%% file: 05_Challenges.tex
\section{Results and Application}
As expected, our results show a diverse range of data-based applications as well as several types of data used within those applications. Especially within Automated Driving and Advanced Driver Assistance Systems, data-driven methods have become a mainstay of research and development. With this, we derived a taxonomy that attempts to classify those data sources and applications into categories that are helpful to stakeholders within organizations. It serves as a basis for an extensible vocabulary focusing on key characteristics that are important within a data-driven development workflow. Attributing the source context of a dataset helps to facilitate reuse of data within a single organization or spanning the whole research community. 
This system of attribution can also help identify gaps in data collected and methods applied within a development process.
\subsection{Data Collection}
Data is sourced in different ways, though real-world data is prevalent within the reviewed literature. We see a focus on physics-related data and with much research being conducted on ADAS and autonomous driving, a lot of data is concerned with the vehicle environment. 
Currently, research relies on a limited number of datasets, often drawn from single sources under specific conditions \cite{Geiger2012CVPR, cordtsCityscapesDatasetSemantic2016, geyerA2D2AudiAutonomous2020}. While some datasets are more diverse, they remain confined to narrowly defined sources or sensors \cite{yuBDD100KDiverseDriving2020}. This lack of coverage becomes problematic when data is used to extract scenarios for requirements elicitation, testing, or training machine learning models. The quality, balance, and comprehensiveness of datasets significantly affect the outcomes of these development tasks \cite{petersen_towards_2022, beringhoff_thirty-one_2022, zhang_finding_2021}. Using our taxonomy, researchers and developers can label datasets and evaluate, whether the selected data is suited to a specific application. 
\subsection{Data-Driven Applications}
The main applications within the development are the modeling and identification of systems or processes, the training of machine learning systems, and verifying and validating those systems. Requirement analysis and conceptual work are a lot less present when considering data-driven research. Current research is focused on machine learning approaches, while classical methods of analytical nature are still relevant. Those surpass the usage of other methods such as probabilistic and rule based approaches or the application of expert knowledge. The study also provides a viewpoint towards different data sources, that can be used to compare and enhance them. Real-world data often consists of diverse data points and has a high realism, while synthetic data may be better at representing a whole parameter space of application conditions \cite{Liu2024}. It is thus useful to use a systematic approach for comparing them in order to decide which dataset or class of datasets is most applicable and practical in certain situations.
\subsection{Application in Data Management}
As previously discussed, a common understanding of business data that relies on a business glossary is necessary in order to promote organizational maturity with respect to CMMI DMM. To this end, we designed our taxonomy as a basis for establishing an extensible business glossary, that is informed by current trends in automotive data-driven methods. It can be used by organizations to form a common understanding of the data that is being generated as well as for selecting suitable applications. With further refinement, the terms within the taxonomy can be used as metadata tags attached to datasets and enable researchers and developers to access data that is fit for their needs. On one hand, it can promote the understanding of the later application at the point of data generation, as well as to establish processes for structured request from users towards data collection. For example, during the early stages of development, a developer can request data that has been marked as synthetic, as it is available in large quantities and can also be tailored to specific scenarios. As validation and verification become more important during later stages of development, more emphasis is put on retrieving real world data in order to focus on a relation to upcoming real world application. 
\if
\else
\begin{itemize}
    \item general framework/platform. Multi purpose Methods and NN Architecture, \cite{papatheocharous_towards_2018}, proposal solution \cite{pillmann_novel_2017} 
    \item Privacy concerns
    \item unbalanced dataset (KITTI, bdd100k,oxford, cityscape,)
    \item security attack \cite{guo_secure_2017}
    \item Data Transfer and Storage 
    \item scenario elicitation for Requirements analysis,  balanced data set as training data \cite{petersen_towards_2022, beringhoff_thirty-one_2022}
    \item Data-Driven System Design \cite{zhang_finding_2021}
    \item Von Korrelation zu Kausalität
\end{itemize}
\fi
\section{Conclusion}
In our work, we reviewed the current literature on automotive data collection. We established a taxonomy that describes the sources and applications of data within the automotive domain.
The proposed taxonomy, which has been derived from current research trends, provides a structured approach for organizing and analyzing automotive data. It provides a common ground for a business glossary that is necessary to reach a mature data management in an organization. In addition, it facilitates the identification of patterns, trends, and insights that can inform decision making in areas such as vehicle design, hardware architecture, and software and update development. 

Our contribution forms part of a future data-centric process in which datasets are collected from different sources throughout an organization. These datasets are enriched with metadata and domain-specific knowledge according to an organization-specific taxonomy. The central data management then receives the data and stores it for future use. The annotations are used to facilitate data quality assurance and the reuse of data by different stakeholders.

To further improve this approach, future research could refine the taxonomy by increasing its granularity and adding more domain specific information such as type of collected physical data, sensor modality, network type or scenario related information such as environment context. Our proposed application categories can be extended, e.g. by classifying  different machine learning approaches or further specifying activities during development. This increases the utility of labeling collections of datasets. Identifying correlations between different elements of the taxonomy (e.g. source types and application patterns) may reveal imbalances in datasets or gaps in data collection, enabling or supporting more complete and representative automotive datasets.

%% file: structuring_data.bbl
\begin{thebibliography}{100}

\bibitem{cmmiinstituteDataManagementMaturity2019}
{CMMI Institute}, ``Data {{Management Maturity}} ({{DMM}}) {{Model
  At-A-Glance}}.''

\bibitem{ALVAREZCOELLO2021}
D.~Alvarez-Coello, D.~Wilms, A.~Bekan, and J.~{Marx Gómez}, ``Towards a
  data-centric architecture in the automotive industry,'' {\em Procedia
  Computer Science}, vol.~181, 2021.
\newblock CENTERIS 2020 - International Conference on ENTERprise Information
  Systems / ProjMAN 2020 - International Conference on Project MANagement /
  HCist 2020 - International Conference on Health and Social Care Information
  Systems and Technologies 2020, CENTERIS/ProjMAN/HCist 2020.

\bibitem{EURECOM_2018}
B.~Klotz, R.~Troncy, D.~Wilms, and C.~Bonnet, ``Vsso - a vehicle signal and
  attribute ontology,'' in {\em SSN 2018, 9th International Semantic Sensor
  Networks Workshop, 9 October 2018, Monterey, CA, USA} (CEUR, ed.),
  (Monterey), 2018.
\newblock CEUR.

\bibitem{Wilms2021}
D.~Wilms, D.~Alvarez-Coello, and A.~Bekan, ``An evolving ontology for vehicle
  signals,'' in {\em 2021 IEEE 93rd Vehicular Technology Conference
  (VTC2021-Spring)}, 2021.

\bibitem{bach_data-driven_2017}
J.~Bach, J.~Langner, S.~Otten, M.~Holzäpfel, and E.~Sax, ``Data-driven
  development, a complementing approach for automotive systems engineering,''
  in {\em 2017 {IEEE} {International} {Systems} {Engineering} {Symposium}
  ({ISSE})}, Oct. 2017.

\bibitem{zhang_finding_2021}
X.~Zhang, J.~Tao, K.~Tan, M.~Törngren, J.~M. Gaspar~Sánchez, M.~Ramli,
  X.~Tao, M.~Gyllenhammar, F.~Wotawa, N.~Mohan, M.~Nica, and H.~Felbinger, {\em
  Finding {Critical} {Scenarios} for {Automated} {Driving} {Systems}: {A}
  {Systematic} {Literature} {Review}}.
\newblock Oct. 2021.

\bibitem{Sohn24}
T.~S. Sohn, M.~Dillitzer, T.~Brüh, R.~Schwager, T.~D. Eberhardt, P.~Elspas,
  and E.~Sax, ``The why, where, when, what and how of data for data-driven
  engineering of automotive systems,'' in {\em 2024 IEEE International
  Symposium on Systems Engineering (ISSE)}, 2024.

\bibitem{gao19}
L.~Gao, L.~Xiong, X.~Lin, X.~Xia, W.~Liu, Y.~Lu, and Z.~Yu, ``Multi-sensor
  {Fusion} {Road} {Friction} {Coefficient} {Estimation} {During} {Steering}
  with {Lyapunov} {Method},'' {\em Sensors}, vol.~19, Jan. 2019.
\newblock 58.

\bibitem{shi_memory-based_2022}
L.~Shi, Z.-H. Zhan, D.~Liang, and J.~Zhang, ``Memory-{Based} {Ant} {Colony}
  {System} {Approach} for {Multi}-{Source} {Data} {Associated} {Dynamic}
  {Electric} {Vehicle} {Dispatch} {Optimization},'' {\em IEEE Transactions on
  Intelligent Transportation Systems}, vol.~23, Oct. 2022.
\newblock 56.

\bibitem{bedretchuk_low-cost_2023}
J.~P. Bedretchuk, S.~Arribas~García, T.~Nogiri~Igarashi, R.~Canal,
  A.~Wedderhoff~Spengler, and G.~Gracioli, ``Low-{Cost} {Data} {Acquisition}
  {System} for {Automotive} {Electronic} {Control} {Units},'' {\em Sensors},
  vol.~23, Jan. 2023.
\newblock 55.

\bibitem{moreira-matias_developing_2017}
L.~Moreira-Matias and H.~Farah, ``On {Developing} a {Driver} {Identification}
  {Methodology} {Using} {In}-{Vehicle} {Data} {Recorders},'' {\em IEEE
  Transactions on Intelligent Transportation Systems}, vol.~18, Sept. 2017.
\newblock 59.

\bibitem{Deo_looking_2018}
M.~M.~T. Nachiket~Deo, ``Looking at the {Driver}/{Rider} in {Autonomous}
  {Vehicles} to {Predict} {Take}-{Over} {Readiness},'' 2018.
\newblock 54.

\bibitem{pham3DDatasetAutonomous2020}
Q.-H. Pham, P.~Sevestre, R.~S. Pahwa, H.~Zhan, C.~H. Pang, Y.~Chen, A.~Mustafa,
  V.~Chandrasekhar, and J.~Lin, ``A*{{3D Dataset}}: {{Towards Autonomous
  Driving}} in {{Challenging Environments}},'' in {\em 2020 {{IEEE
  International Conference}} on {{Robotics}} and {{Automation}} ({{ICRA}})},
  (Paris, France), pp.~2267--2273, IEEE, May 2020.

\bibitem{schmidDataDrivenFaultDiagnosis2021}
M.~Schmid, H.-G. Kneidinger, and C.~Endisch, ``Data-{{Driven Fault Diagnosis}}
  in {{Battery Systems Through Cross-Cell Monitoring}},'' {\em IEEE Sensors
  Journal}, vol.~21, pp.~1829--1837, Jan. 2021.

\bibitem{xingChargingDemandForecasting2019}
Q.~Xing, Z.~Chen, Z.~Zhang, X.~Huang, Z.~Leng, K.~Sun, Y.~Chen, and H.~Wang,
  ``Charging {{Demand Forecasting Model}} for {{Electric Vehicles Based}} on
  {{Online Ride-Hailing Trip Data}},'' {\em IEEE Access}, vol.~7,
  pp.~137390--137409, 2019.

\bibitem{xingDynamicStateEstimation2020}
Y.~Xing and C.~Lv, ``Dynamic {{State Estimation}} for the {{Advanced Brake
  System}} of {{Electric Vehicles}} by {{Using Deep Recurrent Neural
  Networks}},'' {\em IEEE Transactions on Industrial Electronics}, vol.~67,
  pp.~9536--9547, Nov. 2020.

\bibitem{zhaoAcceleratedEvaluationAutomated2018}
D.~Zhao, X.~Huang, H.~Peng, H.~Lam, and D.~J. LeBlanc, ``Accelerated
  {{Evaluation}} of {{Automated Vehicles}} in {{Car-Following Maneuvers}},''
  {\em IEEE Transactions on Intelligent Transportation Systems}, vol.~19,
  pp.~733--744, Mar. 2018.

\bibitem{aminiCabinBatteryThermal2020}
M.~R. Amini, H.~Wang, X.~Gong, D.~{Liao-McPherson}, I.~Kolmanovsky, and J.~Sun,
  ``Cabin and {{Battery Thermal Management}} of {{Connected}} and {{Automated
  HEVs}} for {{Improved Energy Efficiency Using Hierarchical Model Predictive
  Control}},'' {\em IEEE Transactions on Control Systems Technology}, vol.~28,
  pp.~1711--1726, Sept. 2020.

\bibitem{birchlerCosteffectiveSimulationbasedTest2022}
C.~Birchler, N.~Ganz, S.~Khatiri, A.~Gambi, and S.~Panichella, ``Cost-effective
  {{Simulation-based Test Selection}} in {{Self-driving Cars Software}} with
  {{SDC-Scissor}},'' in {\em 2022 {{IEEE International Conference}} on
  {{Software Analysis}}, {{Evolution}} and {{Reengineering}} ({{SANER}})},
  pp.~164--168, Mar. 2022.

\bibitem{degelderScenarioParameterGeneration2022}
E.~De~Gelder, J.~Hof, E.~Cator, J.-P. Paardekooper, O.~O.~D. Camp, J.~Ploeg,
  and B.~De~Schutter, ``Scenario {{Parameter Generation Method}} and {{Scenario
  Representativeness Metric}} for {{Scenario-Based Assessment}} of {{Automated
  Vehicles}},'' {\em IEEE Transactions on Intelligent Transportation Systems},
  vol.~23, pp.~18794--18807, Oct. 2022.

\bibitem{deoLookingDriverRider2020}
N.~Deo and M.~M. Trivedi, ``Looking at the {{Driver}}/{{Rider}} in {{Autonomous
  Vehicles}} to {{Predict Take-Over Readiness}},'' {\em IEEE Transactions on
  Intelligent Vehicles}, vol.~5, pp.~41--52, Mar. 2020.

\bibitem{fowlerFuzzTestingAutomotive2018}
D.~S. Fowler, J.~Bryans, S.~A. Shaikh, and P.~Wooderson, ``Fuzz {{Testing}} for
  {{Automotive Cyber-Security}},'' in {\em 2018 48th {{Annual IEEE}}/{{IFIP
  International Conference}} on {{Dependable Systems}} and {{Networks
  Workshops}} ({{DSN-W}})}, pp.~239--246, June 2018.

\bibitem{gaoPersonalizedAdaptiveCruise2020}
B.~Gao, K.~Cai, T.~Qu, Y.~Hu, and H.~Chen, ``Personalized {{Adaptive Cruise
  Control Based}} on {{Online Driving Style Recognition Technology}} and
  {{Model Predictive Control}},'' {\em IEEE Transactions on Vehicular
  Technology}, vol.~69, pp.~12482--12496, Nov. 2020.

\bibitem{gwonGenerationPreciseEfficient2017}
G.-P. Gwon, W.-S. Hur, S.-W. Kim, and S.-W. Seo, ``Generation of a {{Precise}}
  and {{Efficient Lane-Level Road Map}} for {{Intelligent Vehicle Systems}},''
  {\em IEEE Transactions on Vehicular Technology}, vol.~66, pp.~4517--4533,
  June 2017.

\bibitem{hanResearchRoadEnvironmental2023}
Y.~Han, B.~Wang, T.~Guan, D.~Tian, G.~Yang, W.~Wei, H.~Tang, and J.~H. Chuah,
  ``Research on {{Road Environmental Sense Method}} of {{Intelligent Vehicle
  Based}} on {{Tracking Check}},'' {\em IEEE Transactions on Intelligent
  Transportation Systems}, vol.~24, pp.~1261--1275, Jan. 2023.

\bibitem{jaguemontCharacterizationModelingHybridElectricVehicle2016}
J.~Jaguemont, L.~Boulon, and Y.~Dub{\'e}, ``Characterization and {{Modeling}}
  of a {{Hybrid-Electric-Vehicle Lithium-Ion Battery Pack}} at {{Low
  Temperatures}},'' {\em IEEE Transactions on Vehicular Technology}, vol.~65,
  pp.~1--14, Jan. 2016.

\bibitem{kimTakeoverRequestsSimulated2017}
H.~J. Kim and J.~H. Yang, ``Takeover {{Requests}} in {{Simulated Partially
  Autonomous Vehicles Considering Human Factors}},'' {\em IEEE Transactions on
  Human-Machine Systems}, vol.~47, pp.~735--740, Oct. 2017.

\bibitem{kongFedVCPFederatedLearningBasedCooperative2022}
X.~Kong, H.~Gao, G.~Shen, G.~Duan, and S.~K. Das, ``{{FedVCP}}: {{A
  Federated-Learning-Based Cooperative Positioning Scheme}} for {{Social
  Internet}} of {{Vehicles}},'' {\em IEEE Transactions on Computational Social
  Systems}, vol.~9, pp.~197--206, Feb. 2022.

\bibitem{lvCharacterizationDriverNeuromuscular2018}
C.~Lv, H.~Wang, D.~Cao, Y.~Zhao, D.~J. Auger, M.~Sullman, R.~Matthias,
  L.~Skrypchuk, and A.~Mouzakitis, ``Characterization of {{Driver Neuromuscular
  Dynamics}} for {{Human}}--{{Automation Collaboration Design}} of {{Automated
  Vehicles}},'' {\em IEEE/ASME Transactions on Mechatronics}, vol.~23,
  pp.~2558--2567, Dec. 2018.

\bibitem{lyuUsingNaturalisticDriving2022}
N.~Lyu, Y.~Wang, C.~Wu, L.~Peng, and A.~F. Thomas, ``Using naturalistic driving
  data to identify driving style based on longitudinal driving operation
  conditions,'' {\em Journal of Intelligent and Connected Vehicles}, vol.~5,
  no.~1, pp.~17--35, 2022.

\bibitem{rahmatiHelpingAutomatedVehicles2022}
Y.~Rahmati, M.~K. Hosseini, and A.~Talebpour, ``Helping {{Automated Vehicles
  With Left-Turn Maneuvers}}: {{A Game Theory-Based Decision Framework}} for
  {{Conflicting Maneuvers}} at {{Intersections}},'' {\em IEEE Transactions on
  Intelligent Transportation Systems}, vol.~23, pp.~11877--11890, Aug. 2022.

\bibitem{shinVehicleSpeedPrediction2019}
J.~Shin and M.~Sunwoo, ``Vehicle {{Speed Prediction Using}} a {{Markov Chain
  With Speed Constraints}},'' {\em IEEE Transactions on Intelligent
  Transportation Systems}, vol.~20, pp.~3201--3211, Sept. 2019.

\bibitem{stadlerCredibilityAssessmentApproach2022}
C.~Stadler, F.~Montanari, W.~Baron, C.~Sippl, and A.~Djanatliev, ``A
  {{Credibility Assessment Approach}} for {{Scenario-Based Virtual Testing}} of
  {{Automated Driving Functions}},'' {\em IEEE Open Journal of Intelligent
  Transportation Systems}, vol.~3, pp.~45--60, 2022.

\bibitem{tianDeepTestAutomatedTesting2018}
Y.~Tian, K.~Pei, S.~Jana, and B.~Ray, ``{{DeepTest}}: Automated testing of
  deep-neural-network-driven autonomous cars,'' in {\em Proceedings of the 40th
  {{International Conference}} on {{Software Engineering}}}, {{ICSE}} '18, (New
  York, NY, USA), pp.~303--314, Association for Computing Machinery, May 2018.

\bibitem{wangBChargeDataDrivenRealTime2018}
G.~Wang, X.~Xie, F.~Zhang, Y.~Liu, and D.~Zhang, ``{{bCharge}}: {{Data-Driven
  Real-Time Charging Scheduling}} for {{Large-Scale Electric Bus Fleets}},'' in
  {\em 2018 {{IEEE Real-Time Systems Symposium}} ({{RTSS}})}, pp.~45--55, Dec.
  2018.

\bibitem{wangVibrationTheoreticApproachVulnerability2023}
P.~Wang, X.~Wu, and X.~He, ``Vibration-{{Theoretic Approach}} to
  {{Vulnerability Analysis}} of {{Nonlinear Vehicle Platoons}},'' {\em IEEE
  Transactions on Intelligent Transportation Systems}, vol.~24,
  pp.~11334--11344, Oct. 2023.

\bibitem{xiaoTrajDataVehicleTrajectory2020}
Z.~Xiao, F.~Li, R.~Wu, H.~Jiang, Y.~Hu, J.~Ren, C.~Cai, and A.~Iyengar,
  ``{{TrajData}}: {{On Vehicle Trajectory Collection With Commodity
  Plug-and-Play OBU Devices}},'' {\em IEEE Internet of Things Journal}, vol.~7,
  pp.~9066--9079, Sept. 2020.

\bibitem{santaLPWANBasedVehicularMonitoring2019}
J.~Santa, R.~Sanchez-Iborra, P.~Rodriguez-Rey, L.~Bernal-Escobedo, and A.~F.
  Skarmeta, ``{LPWAN}-{Based} {Vehicular} {Monitoring} {Platform} with a
  {Generic} {IP} {Network} {Interface},'' {\em Sensors}, vol.~19, p.~264, Jan.
  2019.
\newblock 53.

\bibitem{leeDesignImplementationEdgeFogCloud2020}
J.~Lee, K.~Lee, A.~Yoo, and C.~Moon, ``Design and {Implementation} of
  {Edge}-{Fog}-{Cloud} {System} through {HD} {Map} {Generation} from {LiDAR}
  {Data} of {Autonomous} {Vehicles},'' {\em Electronics}, vol.~9, p.~2084, Dec.
  2020.
\newblock 50.

\bibitem{shiConstructingFundamentalDiagram2021}
X.~Shi and X.~Li, ``Constructing a fundamental diagram for traffic flow with
  automated vehicles: {Methodology} and demonstration,'' {\em Transportation
  Research Part B: Methodological}, vol.~150, pp.~279--292, Aug. 2021.
\newblock 48.

\bibitem{shinComparativeStudyMarkov2021}
J.~Shin, K.~Yeon, S.~Kim, M.~Sunwoo, and M.~Han, ``Comparative {Study} of
  {Markov} {Chain} {With} {Recurrent} {Neural} {Network} for {Short} {Term}
  {Velocity} {Prediction} {Implemented} on an {Embedded} {System},'' {\em IEEE
  Access}, vol.~9, pp.~24755--24767, 2021.
\newblock 46.

\bibitem{jaguemontCharacterizationModelingHybridElectricVehicle2016a}
J.~Jaguemont, L.~Boulon, and Y.~Dubé, ``Characterization and {Modeling} of a
  {Hybrid}-{Electric}-{Vehicle} {Lithium}-{Ion} {Battery} {Pack} at {Low}
  {Temperatures},'' {\em IEEE Transactions on Vehicular Technology}, vol.~65,
  pp.~1--14, Jan. 2016.
\newblock 45.

\bibitem{leledakisMethodPredictingCrash2021}
A.~Leledakis, M.~Lindman, J.~Östh, L.~Wågström, J.~Davidsson, and
  L.~Jakobsson, ``A method for predicting crash configurations using
  counterfactual simulations and real-world data,'' {\em Accident Analysis \&
  Prevention}, vol.~150, p.~105932, Feb. 2021.
\newblock 41.

\bibitem{choiTrajGAILGeneratingUrban2021}
S.~Choi, J.~Kim, and H.~Yeo, ``{TrajGAIL}: {Generating} urban vehicle
  trajectories using generative adversarial imitation learning,'' {\em
  Transportation Research Part C: Emerging Technologies}, vol.~128, p.~103091,
  July 2021.
\newblock 37.

\bibitem{CredibilityAssessmentApproach}
``A {Credibility} {Assessment} {Approach} for {Scenario}-{Based} {Virtual}
  {Testing} of {Automated} {Driving} {Functions} {\textbar} {IEEE} {Journals}
  \& {Magazine} {\textbar} {IEEE} {Xplore}.''
\newblock 38.

\bibitem{benkraoudaTrafficDataImputation2020}
O.~Benkraouda, B.~T. Thodi, H.~Yeo, M.~Menéndez, and S.~E. Jabari, ``Traffic
  {Data} {Imputation} {Using} {Deep} {Convolutional} {Neural} {Networks},''
  {\em IEEE Access}, vol.~8, pp.~104740--104752, 2020.
\newblock 36.

\bibitem{arvinRolePrecrashDriving2019}
R.~Arvin, M.~Kamrani, and A.~J. Khattak, ``The role of pre-crash driving
  instability in contributing to crash intensity using naturalistic driving
  data,'' {\em Accident Analysis \& Prevention}, vol.~132, p.~105226, Nov.
  2019.
\newblock 35.

\bibitem{caiIntegrationGPSMonocular2018}
H.~Cai, Z.~Hu, G.~Huang, D.~Zhu, and X.~Su, ``Integration of {GPS}, {Monocular}
  {Vision}, and {High} {Definition} ({HD}) {Map} for {Accurate} {Vehicle}
  {Localization},'' {\em Sensors}, vol.~18, p.~3270, Oct. 2018.
\newblock 33.

\bibitem{ILOCuSIncentivizingVehicle}
``{iLOCuS}: {Incentivizing} {Vehicle} {Mobility} to {Optimize} {Sensing}
  {Distribution} in {Crowd} {Sensing} {\textbar} {IEEE} {Journals} \&
  {Magazine} {\textbar} {IEEE} {Xplore}.''
\newblock 32.

\bibitem{waliHowDrivingVolatility2018}
B.~Wali, A.~J. Khattak, H.~Bozdogan, and M.~Kamrani, ``How is driving
  volatility related to intersection safety? {A} {Bayesian} heterogeneity-based
  analysis of instrumented vehicles data,'' {\em Transportation Research Part
  C: Emerging Technologies}, vol.~92, pp.~504--524, July 2018.
\newblock 31.

\bibitem{liuDatadrivenEnergyManagement2021}
J.~Liu, Z.~Wang, Y.~Hou, C.~Qu, J.~Hong, and N.~Lin, ``Data-driven energy
  management and velocity prediction for four-wheel-independent-driving
  electric vehicles,'' {\em eTransportation}, vol.~9, p.~100119, Aug. 2021.
\newblock 27.

\bibitem{liuCustomizingDrivingCycles2016}
J.~Liu, X.~Wang, and A.~Khattak, ``Customizing driving cycles to support
  vehicle purchase and use decisions: {Fuel} economy estimation for alternative
  fuel vehicle users,'' {\em Transportation Research Part C: Emerging
  Technologies}, vol.~67, pp.~280--298, June 2016.
\newblock 26.

\bibitem{xingChargingDemandForecasting2019a}
Q.~Xing, Z.~Chen, Z.~Zhang, X.~Huang, Z.~Leng, K.~Sun, Y.~Chen, and H.~Wang,
  ``Charging {Demand} {Forecasting} {Model} for {Electric} {Vehicles} {Based}
  on {Online} {Ride}-{Hailing} {Trip} {Data},'' {\em IEEE Access}, vol.~7,
  pp.~137390--137409, 2019.
\newblock 24.

\bibitem{zhaoTrafficNetOpenNaturalistic2017}
D.~Zhao, Y.~Guo, and Y.~J. Jia, ``{TrafficNet}: {An} open naturalistic driving
  scenario library,'' in {\em 2017 {IEEE} 20th {International} {Conference} on
  {Intelligent} {Transportation} {Systems} ({ITSC})}, pp.~1--8, Oct. 2017.
\newblock 22.

\bibitem{zhaoElectricVehicleBatteries2022}
J.~Zhao and A.~F. Burke, ``Electric {Vehicle} {Batteries}: {Status} and
  {Perspectives} of {Data}-{Driven} {Diagnosis} and {Prognosis},'' {\em
  Batteries}, vol.~8, p.~142, Oct. 2022.
\newblock 20.

\bibitem{vicenteLinearSystemIdentification2021}
B.~A.~H. Vicente, S.~S. James, and S.~R. Anderson, ``Linear {System}
  {Identification} {Versus} {Physical} {Modeling} of {Lateral}–{Longitudinal}
  {Vehicle} {Dynamics},'' {\em IEEE Transactions on Control Systems
  Technology}, vol.~29, pp.~1380--1387, May 2021.
\newblock 11.

\bibitem{loniDatadrivenEquitablePlacement2023}
A.~Loni and S.~Asadi, ``Data-driven equitable placement for electric vehicle
  charging stations: {Case} study {San} {Francisco},'' {\em Energy}, vol.~282,
  p.~128796, Nov. 2023.
\newblock 17.

\bibitem{haqueLoRaArchitectureV2X2020}
K.~F. Haque, A.~Abdelgawad, V.~P. Yanambaka, and K.~Yelamarthi, ``{LoRa}
  {Architecture} for {V2X} {Communication}: {An} {Experimental} {Evaluation}
  with {Vehicles} on the {Move},'' {\em Sensors}, vol.~20, p.~6876, Jan. 2020.
\newblock 12.

\bibitem{fridmanMITAdvancedVehicle2019b}
L.~Fridman, D.~E. Brown, M.~Glazer, W.~Angell, S.~Dodd, B.~Jenik,
  J.~Terwilliger, A.~Patsekin, J.~Kindelsberger, L.~Ding, S.~Seaman, A.~Mehler,
  A.~Sipperley, A.~Pettinato, B.~D. Seppelt, L.~Angell, B.~Mehler, and
  B.~Reimer, ``{MIT} {Advanced} {Vehicle} {Technology} {Study}: {Large}-{Scale}
  {Naturalistic} {Driving} {Study} of {Driver} {Behavior} and {Interaction}
  {With} {Automation},'' {\em IEEE Access}, vol.~7, pp.~102021--102038, 2019.
\newblock 2.

\bibitem{chenAutonomousVehicleTesting2018}
Y.~Chen, S.~Chen, T.~Zhang, S.~Zhang, and N.~Zheng, ``Autonomous {{Vehicle
  Testing}} and {{Validation Platform}}: {{Integrated Simulation System}} with
  {{Hardware}} in the {{Loop}},'' in {\em 2018 {{IEEE Intelligent Vehicles
  Symposium}} ({{IV}})}, pp.~949--956, June 2018.

\bibitem{chenDeepImitationLearning2019}
J.~Chen, B.~Yuan, and M.~Tomizuka, ``Deep {{Imitation Learning}} for
  {{Autonomous Driving}} in {{Generic Urban Scenarios}} with {{Enhanced
  Safety}},'' in {\em 2019 {{IEEE}}/{{RSJ International Conference}} on
  {{Intelligent Robots}} and {{Systems}} ({{IROS}})}, pp.~2884--2890, Nov.
  2019.

\bibitem{chenRHONNModellingEnabledNonlinear2022}
H.~Chen, J.~Zhang, and C.~Lv, ``{{RHONN Modelling-Enabled Nonlinear Predictive
  Control}} for {{Lateral Dynamics Stabilization}} of an {{In-Wheel Motor
  Driven Vehicle}},'' {\em IEEE Transactions on Vehicular Technology}, vol.~71,
  pp.~8296--8308, Aug. 2022.

\bibitem{gidadoSurveyDeepLearning2020}
U.~M. Gidado, H.~Chiroma, N.~Aljojo, S.~Abubakar, S.~I. Popoola, and M.~A.
  {Al-Garadi}, ``A {{Survey}} on {{Deep Learning}} for {{Steering Angle
  Prediction}} in {{Autonomous Vehicles}},'' {\em IEEE Access}, vol.~8,
  pp.~163797--163817, 2020.

\bibitem{luLearningDriverSpecificBehavior2018}
C.~Lu, H.~Wang, C.~Lv, J.~Gong, J.~Xi, and D.~Cao, ``Learning {{Driver-Specific
  Behavior}} for {{Overtaking}}: {{A Combined Learning Framework}},'' {\em IEEE
  Transactions on Vehicular Technology}, vol.~67, pp.~6788--6802, Aug. 2018.

\bibitem{qiLearningBasedPathPlanning2023}
Z.~Qi, T.~Wang, J.~Chen, D.~Narang, Y.~Wang, and H.~Yang, ``Learning-{{Based
  Path Planning}} and {{Predictive Control}} for {{Autonomous Vehicles With
  Low-Cost Positioning}},'' {\em IEEE Transactions on Intelligent Vehicles},
  vol.~8, pp.~1093--1104, Feb. 2023.

\bibitem{tangATACBasedCarFollowingModel2022}
T.-Q. Tang, Y.~Gui, and J.~Zhang, ``{{ATAC-Based Car-Following Model}} for
  {{Level}} 3 {{Autonomous Driving Considering Driver}}'s {{Acceptance}},''
  {\em IEEE Transactions on Intelligent Transportation Systems}, vol.~23,
  pp.~10309--10321, Aug. 2022.

\bibitem{wangSFNetNImprovedSFNet2022}
H.~Wang, Y.~Chen, Y.~Cai, L.~Chen, Y.~Li, M.~A. Sotelo, and Z.~Li,
  ``{{SFNet-N}}: {{An Improved SFNet Algorithm}} for {{Semantic Segmentation}}
  of {{Low-Light Autonomous Driving Road Scenes}},'' {\em IEEE Transactions on
  Intelligent Transportation Systems}, vol.~23, pp.~21405--21417, Nov. 2022.

\bibitem{weiContractBasedChargingProtocol2023}
Z.~Wei, B.~Li, R.~Zhang, and X.~Cheng, ``Contract-{{Based Charging Protocol}}
  for {{Electric Vehicles With Vehicular Fog Computing}}: {{An Integrated
  Charging}} and {{Computing Perspective}},'' {\em IEEE Internet of Things
  Journal}, vol.~10, pp.~7667--7680, May 2023.

\bibitem{xuGenerativeAIEmpoweredSimulation2023}
M.~Xu, D.~Niyato, J.~Chen, H.~Zhang, J.~Kang, Z.~Xiong, S.~Mao, and Z.~Han,
  ``Generative {{AI-Empowered Simulation}} for {{Autonomous Driving}} in
  {{Vehicular Mixed Reality Metaverses}},'' {\em IEEE Journal of Selected
  Topics in Signal Processing}, vol.~17, pp.~1064--1079, Sept. 2023.

\bibitem{wangDataDrivenPredictiveControl2022}
J.~Wang, Y.~Zheng, Q.~Xu, and K.~Li, ``Data-{Driven} {Predictive} {Control} for
  {Connected} and {Autonomous} {Vehicles} in {Mixed} {Traffic},'' in {\em 2022
  {American} {Control} {Conference} ({ACC})}, pp.~4739--4745, June 2022.
\newblock 49.

\bibitem{nieDeepNeuralNetworkBasedModellingLongitudinalLateral2022}
X.~Nie, C.~Min, Y.~Pan, K.~Li, and Z.~Li, ``Deep-{Neural}-{Network}-{Based}
  {Modelling} of {Longitudinal}-{Lateral} {Dynamics} to {Predict} the {Vehicle}
  {States} for {Autonomous} {Driving},'' {\em Sensors}, vol.~22, p.~2013, Jan.
  2022.
\newblock 28.

\bibitem{cuiCoopernautEndtoEndDriving2022}
J.~Cui, H.~Qiu, D.~Chen, P.~Stone, and Y.~Zhu, ``Coopernaut: {End}-to-{End}
  {Driving} with {Cooperative} {Perception} for {Networked} {Vehicles},'' in
  {\em 2022 {IEEE}/{CVF} {Conference} on {Computer} {Vision} and {Pattern}
  {Recognition} ({CVPR})}, pp.~17231--17241, June 2022.
\newblock 16.

\bibitem{tangDriverLaneChange2020a}
L.~Tang, H.~Wang, W.~Zhang, Z.~Mei, and L.~Li, ``Driver {Lane} {Change}
  {Intention} {Recognition} of {Intelligent} {Vehicle} {Based} on {Long}
  {Short}-{Term} {Memory} {Network},'' {\em IEEE Access}, vol.~8,
  pp.~136898--136905, 2020.
\newblock 5.

\bibitem{berziDevelopmentDrivingCycles2016a}
L.~Berzi, M.~Delogu, and M.~Pierini, ``Development of driving cycles for
  electric vehicles in the context of the city of {Florence},'' {\em
  Transportation Research Part D: Transport and Environment}, vol.~47,
  pp.~299--322, Aug. 2016.
\newblock 7.

\bibitem{weiEfficientDatadrivenOptimal2022}
Y.~Wei, T.~Han, S.~Wang, Y.~Qin, L.~Lu, X.~Han, and M.~Ouyang, ``An efficient
  data-driven optimal sizing framework for photovoltaics-battery-based electric
  vehicle charging microgrid,'' {\em Journal of Energy Storage}, vol.~55,
  p.~105670, Nov. 2022.
\newblock 15.

\bibitem{ebelForcesDriverDistraction2023a}
P.~Ebel, C.~Lingenfelder, and A.~Vogelsang, ``On the forces of driver
  distraction: {Explainable} predictions for the visual demand of in-vehicle
  touchscreen interactions,'' {\em Accident Analysis \& Prevention}, vol.~183,
  p.~106956, Apr. 2023.
\newblock 8.

\bibitem{fengDesignDistributedCyber2018}
Y.~Feng, B.~Hu, H.~Hao, Y.~Gao, Z.~Li, and J.~Tan, ``Design of {Distributed}
  {Cyber}–{Physical} {Systems} for {Connected} and {Automated} {Vehicles}
  {With} {Implementing} {Methodologies},'' {\em IEEE Transactions on Industrial
  Informatics}, vol.~14, pp.~4200--4211, Sept. 2018.
\newblock 51.

\bibitem{deyVehicletovehicleV2VVehicletoinfrastructure2016}
K.~C. Dey, A.~Rayamajhi, M.~Chowdhury, P.~Bhavsar, and J.~Martin,
  ``Vehicle-to-vehicle ({V2V}) and vehicle-to-infrastructure ({V2I})
  communication in a heterogeneous wireless network – {Performance}
  evaluation,'' {\em Transportation Research Part C: Emerging Technologies},
  vol.~68, pp.~168--184, July 2016.
\newblock 13.

\bibitem{moreira-matiasDevelopingDriverIdentification2017}
L.~Moreira-Matias and H.~Farah, ``On {Developing} a {Driver} {Identification}
  {Methodology} {Using} {In}-{Vehicle} {Data} {Recorders},'' {\em IEEE
  Transactions on Intelligent Transportation Systems}, vol.~18, pp.~2387--2396,
  Sept. 2017.
\newblock 59.

\bibitem{neurohrCriticalityAnalysisVerification2021}
C.~Neurohr, L.~Westhofen, M.~Butz, M.~H. Bollmann, U.~Eberle, and R.~Galbas,
  ``Criticality {Analysis} for the {Verification} and {Validation} of
  {Automated} {Vehicles},'' {\em IEEE Access}, vol.~9, pp.~18016--18041, 2021.
\newblock 25.

\bibitem{aliMachineLearningTechnologies2021}
E.~S. Ali, M.~K. Hasan, R.~Hassan, R.~A. Saeed, M.~B. Hassan, S.~Islam, N.~S.
  Nafi, and S.~Bevinakoppa, ``Machine {Learning} {Technologies} for {Secure}
  {Vehicular} {Communication} in {Internet} of {Vehicles}: {Recent} {Advances}
  and {Applications},'' {\em Security and Communication Networks}, vol.~2021,
  no.~1, p.~8868355, 2021.
\newblock 34.

\bibitem{bedretchukLowCostDataAcquisition2023}
J.~P. Bedretchuk, S.~Arribas~García, T.~Nogiri~Igarashi, R.~Canal,
  A.~Wedderhoff~Spengler, and G.~Gracioli, ``Low-{Cost} {Data} {Acquisition}
  {System} for {Automotive} {Electronic} {Control} {Units},'' {\em Sensors},
  vol.~23, p.~2319, Jan. 2023.
\newblock 55.

\bibitem{DesignTrafficEmergency}
``Design of {Traffic} {Emergency} {Response} {System} {Based} on {Internet} of
  {Things} and {Data} {Mining} in {Emergencies}.''
\newblock 52.

\bibitem{LookingDriverRider}
``Looking at the {Driver}/{Rider} in {Autonomous} {Vehicles} to {Predict}
  {Take}-{Over} {Readiness}.''
\newblock 54.

\bibitem{ostadianIntelligentEnergyManagement2020a}
R.~Ostadian, J.~Ramoul, A.~Biswas, and A.~Emadi, ``Intelligent {Energy}
  {Management} {Systems} for {Electrified} {Vehicles}: {Current} {Status},
  {Challenges}, and {Emerging} {Trends},'' {\em IEEE Open Journal of Vehicular
  Technology}, vol.~1, pp.~279--295, 2020.
\newblock 6.

\bibitem{zouVehicleAccelerationPrediction2022}
Y.~Zou, L.~Ding, H.~Zhang, T.~Zhu, and L.~Wu, ``Vehicle {Acceleration}
  {Prediction} {Based} on {Machine} {Learning} {Models} and {Driving}
  {Behavior} {Analysis},'' {\em Applied Sciences}, vol.~12, p.~5259, Jan. 2022.
\newblock 64.

\bibitem{huangMultitargetPredictionOptimization2023b}
H.~Huang, T.~C. Lim, J.~Wu, W.~Ding, and J.~Pang, ``Multitarget prediction and
  optimization of pure electric vehicle tire/road airborne noise sound quality
  based on a knowledge- and data-driven method,'' {\em Mechanical Systems and
  Signal Processing}, vol.~197, p.~110361, Aug. 2023.

\bibitem{jiaPlatoonBasedCooperative2016}
D.~Jia and D.~Ngoduy, ``Platoon based cooperative driving model with
  consideration of realistic inter-vehicle communication,'' {\em Transportation
  Research Part C: Emerging Technologies}, vol.~68, pp.~245--264, July 2016.
\newblock 60.

\bibitem{qiTrafficDifferentiatedClustering2020}
W.~Qi, B.~Landfeldt, Q.~Song, L.~Guo, and A.~Jamalipour, ``Traffic
  {Differentiated} {Clustering} {Routing} in {DSRC} and {C}-{V2X} {Hybrid}
  {Vehicular} {Networks},'' {\em IEEE Transactions on Vehicular Technology},
  vol.~69, pp.~7723--7734, July 2020.
\newblock 63.

\bibitem{gaoMultisensorFusionRoad2019}
L.~Gao, L.~Xiong, X.~Lin, X.~Xia, W.~Liu, Y.~Lu, and Z.~Yu, ``Multi-sensor
  {Fusion} {Road} {Friction} {Coefficient} {Estimation} {During} {Steering}
  with {Lyapunov} {Method},'' {\em Sensors}, vol.~19, p.~3816, Jan. 2019.
\newblock 58.

\bibitem{yaoDynamicPredictiveTraffic2020}
Z.~Yao, L.~Shen, R.~Liu, Y.~Jiang, and X.~Yang, ``A {Dynamic} {Predictive}
  {Traffic} {Signal} {Control} {Framework} in a {Cross}-{Sectional} {Vehicle}
  {Infrastructure} {Integration} {Environment},'' {\em IEEE Transactions on
  Intelligent Transportation Systems}, vol.~21, pp.~1455--1466, Apr. 2020.
\newblock 39.

\bibitem{almeaibedDigitalTwinAnalysis2021}
S.~Almeaibed, S.~Al-Rubaye, A.~Tsourdos, and N.~P. Avdelidis, ``Digital {Twin}
  {Analysis} to {Promote} {Safety} and {Security} in {Autonomous} {Vehicles},''
  {\em IEEE Communications Standards Magazine}, vol.~5, pp.~40--46, Mar. 2021.
\newblock 19.

\bibitem{cebeBlock4ForensicIntegratedLightweight2018}
M.~Cebe, E.~Erdin, K.~Akkaya, H.~Aksu, and S.~Uluagac, ``{Block4Forensic}: {An}
  {Integrated} {Lightweight} {Blockchain} {Framework} for {Forensics}
  {Applications} of {Connected} {Vehicles},'' {\em IEEE Communications
  Magazine}, vol.~56, pp.~50--57, Oct. 2018.
\newblock 43.

\bibitem{parkScenarioMiningLevel42021a}
S.~Park, S.~Park, H.~Jeong, I.~Yun, and J.~J. So, ``Scenario-{Mining} for
  {Level} 4 {Automated} {Vehicle} {Safety} {Assessment} from {Real} {Accident}
  {Situations} in {Urban} {Areas} {Using} a {Natural} {Language} {Process},''
  {\em Sensors}, vol.~21, p.~6929, Jan. 2021.
\newblock 67.

\bibitem{huang_multitarget_2023}
H.~Huang, T.~C. Lim, J.~Wu, W.~Ding, and J.~Pang, ``Multitarget prediction and
  optimization of pure electric vehicle tire/road airborne noise sound quality
  based on a knowledge- and data-driven method,'' {\em Mechanical Systems and
  Signal Processing}, vol.~197, Aug. 2023.

\bibitem{Geiger2012CVPR}
A.~Geiger, P.~Lenz, and R.~Urtasun, ``Are we ready for autonomous driving? the
  kitti vision benchmark suite,'' in {\em Conference on Computer Vision and
  Pattern Recognition (CVPR)}, 2012.

\bibitem{cordtsCityscapesDatasetSemantic2016}
M.~Cordts, M.~Omran, S.~Ramos, T.~Rehfeld, M.~Enzweiler, R.~Benenson,
  U.~Franke, S.~Roth, and B.~Schiele, ``The {{Cityscapes Dataset}} for
  {{Semantic Urban Scene Understanding}},'' Apr. 2016.

\bibitem{geyerA2D2AudiAutonomous2020}
J.~Geyer, Y.~Kassahun, M.~Mahmudi, X.~Ricou, R.~Durgesh, A.~S. Chung,
  L.~Hauswald, V.~H. Pham, M.~M{\"u}hlegg, S.~Dorn, T.~Fernandez,
  M.~J{\"a}nicke, S.~Mirashi, C.~Savani, M.~Sturm, O.~Vorobiov, M.~Oelker,
  S.~Garreis, and P.~Schuberth, ``{{A2D2}}: {{Audi Autonomous Driving
  Dataset}},'' Apr. 2020.

\bibitem{yuBDD100KDiverseDriving2020}
F.~Yu, H.~Chen, X.~Wang, W.~Xian, Y.~Chen, F.~Liu, V.~Madhavan, and T.~Darrell,
  ``{{BDD100K}}: {{A Diverse Driving Dataset}} for {{Heterogeneous Multitask
  Learning}},'' Apr. 2020.

\bibitem{petersen_towards_2022}
P.~Petersen, H.~Stage, J.~Langner, L.~Ries, P.~Rigoll, C.~Philipp~Hohl, and
  E.~Sax, ``Towards a {Data} {Engineering} {Process} in {Data}-{Driven}
  {Systems} {Engineering},'' in {\em 2022 {IEEE} {International} {Symposium} on
  {Systems} {Engineering} ({ISSE})}, Oct. 2022.
\newblock ISSN: 2687-8828.

\bibitem{beringhoff_thirty-one_2022}
F.~Beringhoff, J.~Greenyer, C.~Roesener, and M.~Tichy, ``Thirty-{One}
  {Challenges} in {Testing} {Automated} {Vehicles}: {Interviews} with {Experts}
  from {Industry} and {Research},'' in {\em 2022 {IEEE} {Intelligent}
  {Vehicles} {Symposium} ({IV})}, June 2022.

\bibitem{Liu2024}
H.~X. Liu and S.~Feng, ``Curse of rarity for autonomous vehicles,'' {\em Nature
  Communications}, vol.~15, p.~4808, Jun 2024.

\end{thebibliography}
